\documentclass[bibnotes, preprint, aps]{revtex4}

\usepackage{color,epsfig,rotating,graphicx,subfigure}
\usepackage{amsmath,amssymb,amscd}

\usepackage[latin1]{inputenc}
\usepackage[english]{babel}

\usepackage{dsfont}
\usepackage{textcomp}

\begin{document}

\newcommand{\be}{\begin{equation}}
\newcommand{\ee}{\end{equation}}
\newcommand{\bea}{\begin{eqnarray}}
\newcommand{\eea}{\end{eqnarray}}
\newcommand{\bwt}{\begin{widetext}}
\newcommand{\ewt}{\end{widetext}}

\newcommand{\ri}{{\rm i}}
\newcommand{\re}{{\rm e}}
\newcommand{\rd}{{\rm d}}
\newcommand{\rr}{{\rm r}}
\newcommand{\rt}{{\rm t}}
\newcommand{\rl}{{\rm l}}  
\newcommand{\rh}{{\rm h}}
\newcommand{\rb}{{\rm b}}
\newcommand{\rs}{{\rm s}}
\newcommand{\rp}{{\rm p}}  
\newcommand{\rth}{{\rm th}}  
\newcommand{\kb}{k_{\rm B}}

\newcommand{\Tr}{{\rm Tr}}

\title{Modulation of near-field heat transfer between two gratings}

\author{S.-A. Biehs} 
\affiliation{Laboratoire Charles Fabry, Institut d'Optique, CNRS, Universit\'{e} Paris-Sud, Campus
Polytechnique, RD128, 91127 Palaiseau Cedex, France}

\author{F.S.S. Rosa}
\affiliation{Laboratoire Charles Fabry, Institut d'Optique, CNRS, Universit\'{e} Paris-Sud, Campus
Polytechnique, RD128, 91127 Palaiseau Cedex, France}

\author{P. Ben-Abdallah}
\affiliation{Laboratoire Charles Fabry, Institut d'Optique, CNRS, Universit\'{e} Paris-Sud, Campus
Polytechnique, RD128, 91127 Palaiseau Cedex, France}

\date{\today}
%\pacs{44.40.+a, 78.20.-e,03.50.De}

\begin{abstract}
We present a theoretical study of near-field heat transfer between two uniaxial anisotropic planar structures. 
We investigate how the distance and relative orientation (with respect to their optical axes) 
between the objects affect the heat flux. In particular, we show that by changing the angle between the optical axes 
it is possible in certain cases to modulate the net heat flux up to 90\% at room temperature, and discuss possible 
applications of such a strong effect.
\end{abstract}

\maketitle

%%%%%%%%%%%%%%%%%%%%%%%%%%%%
%%%%%%%%%%%%%%%%%%%%%%%%%%%%

Since the prediction by Polder and van Hove \cite{Polder1973} that the heat exchange between two media at short separations can be much higher 
than the blackbody limit, numerous works have been carried out to investigate both theoretically and experimentally the physics involved in 
this transfer. Experimentally, it was shown~\cite{Kittel,HuEtAl2008} that the radiative heat flux increases for distances shorter than the thermal wavelength 
and can vastly exceed the black body limit~\cite{NarayaEtAl2008,ShenEtAl2008}. Moreover,  very recent experiments~\cite{RousseauEtAl2009,Ottens2011} were in good quantitative agreement with theoretical predictions. On the theoretical side, we can highlight the studies of the heat flux for layered media~\cite{Biehs2007,PBA2009}, for photonic crystals~\cite{PBA2010}, metamaterials~\cite{Joulain2010}, and porous media~\cite{BiehsEtAl2011}. In addition, the dependence of the heat transfer on the geometry has attracted much interest and 
has been investigated in a sphere-plane geometry~\cite{OteyFan2011,Krueger2011}, for spheroidal particles above a plane surface~\cite{HuthEtAl2010} and between two spheres or nanoparticles~\cite{ChapuisEtAl2008,Nara2008,Domingues2005,Perez2008,Perez2009}. Somewhat more applied studies have attempted to take advantage of the potential of the tremendous increase of 
the radiative heat flux on the nanoscale for thermal imaging of nanostructured surfaces~\cite{BiehsEtAl2008,Kittel2008,Rueting2010,Biehs2010E}. 
Finally, the formulation of the heat flux in terms of the scattering matrix~\cite{Bimonte2009,Messina2010} paves the way for the study of further geometries and 
the Landauer concept~\cite{BiehsEtAl2010,JoulainPBA2010} opens up a deeper understanding of the trade-off between heat transmission and the number of modes contributing to the heat flux.

While considerable progress has been made over the last decades to actively manage heat flow carried by phonons in nanostructures~\cite{ChangEtAl2006,Wang2007,Segal2008}, very few attention 
has been paid so far on the control of non-contact heat exchanges at nanoscale. In 2010, a pioneer work carried out in this way by Otey {\itshape et al.}~\cite{Fan2010} 
proposed a thermal rectifier based on photon tunneling between two thermally dependent polar materials separated by a vacuum gap. The efficiency of the thermal 
rectification was found to be about $40\%$. More recently, a device made with phase change materials has been introduced by van Zwol {\itshape et al.}~\cite{Zwol2010} to modulate
heat flux between two materials using an electric ac current as external power source. However, due to the  properties of these materials, such modulator works 
reversibly only during a limited number of cycle which typically oscillate between $10^7$ and $10^{12}$. Moreover, such devices work at two discrete levels of flux, 
one for each state of the phase changing material. 

In this Letter, we investigate the near-field heat transfer between two polar/metallic misaligned gratings in the long wavelength limit, where they may be described by effective homogeneous anisotropic permittivities. We show that it is possible to get a strong heat flux modulation without cycle limitation just by rotating the relative position of the grating's optical axes \cite{Mccauley10}. Our approach combines the standard stochastic electrodynamics~\cite{Rytov} and the effective medium theory~\cite{TaoEtAl1990,HaggansEtAl1993,BrundrettEtAl1994} for the gratings.

\begin{figure}
\begin{center}
\scalebox{0.65}{\includegraphics{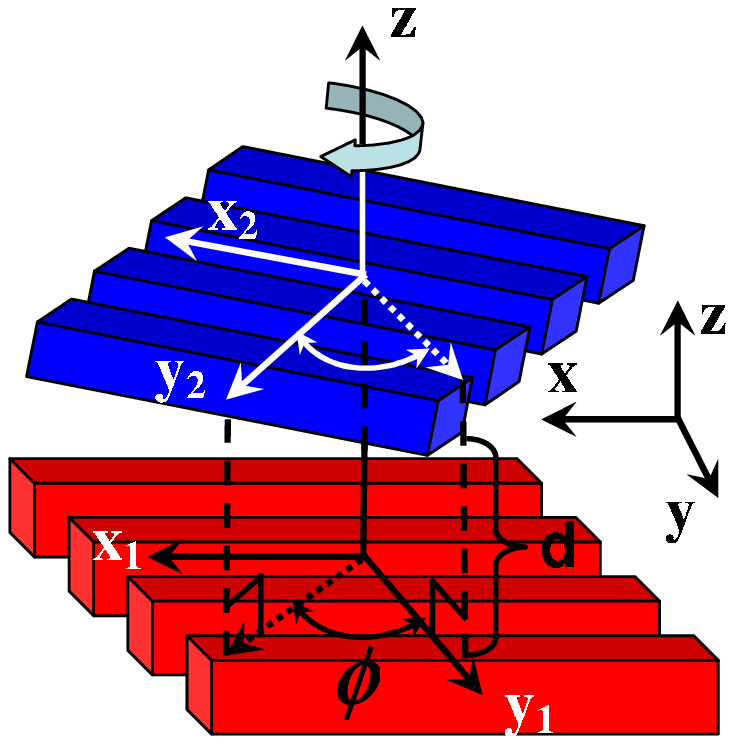}}
\caption{Two gratings separated by a distance d, and relatively twisted by an angle $\phi$.}
\label{Fig1}
\end{center}
\end{figure}

A sketch of the geometry considered is depicted in Fig.~\ref{Fig1}. It shows two semi-infinite host materials of complex permittivity $\epsilon_i(\omega)$ ($i = 1,2$) with a one dimensional grating engraved on each. The relative orientation of the two gratings is arbitrary in the (x,y) plane, and we assume that their trenches are sufficiently deep so as to (i) render the substrate below those gratings irrelevant and (ii) allow us to consider the x- and z-directions as equivalent.  Moreover, these structures are separated by a vacuum gap of thickness d and kept at two different temperatures $T_1$, $T_2$ in local thermal equilibrium. By choosing a coordinate system with the z-axis perpendicular to the interfaces, we can write the permittivity in the form \cite{Thesis, Rosa2008}
\begin{equation}
\bar{\bar{\epsilon}} = \left[ 
\begin{array}{ccc}
\epsilon_{\bot_i} \sin^2\phi_i  +  \epsilon_{\|_i} \cos^2 \phi_i & (\epsilon_{\bot_i}-\epsilon_{\|_i}) \sin\phi_i \cos\phi_i & 0 \\
(\epsilon_{\bot_i}-\epsilon_{\|_i}) \sin\phi_i \cos\phi_i & \epsilon_{\bot_i} \cos^2\phi_i  +  \epsilon_{\|_i} \sin^2 \phi_i &  0 \\
0 &
0 & 
\epsilon_{\|}
\end{array} \right] ,
\end{equation}
where $\phi_i$ is the angle between the y- and the i-th structure's optical axes, and $\epsilon_{\|}$,  $\epsilon_{\bot}$ are given by the effective medium expressions~\cite{TaoEtAl1990,HaggansEtAl1993,BrundrettEtAl1994} 
\be
\epsilon_{\|} = \epsilon_{\rh_i}(1-f_i) + f_i \,\,\,\,\,\, , \,\,\,\,\,\, \epsilon_{\bot_i} = \frac{\epsilon_{\rh_i}}{(1-f_i)+f_i \epsilon_{\rh_i}} ,
\label{Eq:EpsEMT}
\ee
where $f_i$ and $\epsilon_{h_i}$ are respectively the vacuum filling factor and permittivity of the host material in the i-th grating. These expressions are valid for arbitrary filling factors
as long as the grating periods $\Lambda_\ri$ are much smaller than the thermal wavelength $\lambda_\rth = c \hbar / \kb T$, that in our case ($300\,{\rm K}$) is about $7.68\,\mu{\rm m}$.
Within the near-field regime this condition for the validity of the homogenization is different. As discussed in more detail in Ref.~\cite{BiehsEtAl2011} the expressions for the effective permittivity in the near-field regime are valid if $\Lambda_\ri$ is smaller than the spacing $d$ between the gratings.

According to standard stochastic electrodynamics~\cite{Rytov,Polder1973}, the heat flux exchanged between the two bodies per unit surface is given by the statistical average of the Poynting vector normal component $S_z$
\begin{equation}
  \langle S_z \rangle =  \int_0^\infty\!\frac{\rd \omega}{2 \pi} \left[ \Theta(\omega,T_1) - \Theta(\omega,T_2) \right]  \int \frac{d{\bf k}_{\|}}{(2\pi)^2} T(\omega, k_{\|}),
\label{Eq:MeanPoynting}
\end{equation}
where $\Theta(\omega,T) = \hbar \omega / [e^{\hbar \omega / k_B T} - 1]$ is the mean energy of a harmonic oscillator and the transmission factor $T(\omega, k_{\|})$ \cite{BiehsEtAl2010}
can be written as~\cite{BiehsEtAl2011} 
\begin{equation}
\begin{split}
   T(\omega,\boldsymbol{\kappa}; d) = 
    \begin{cases}
     \Tr \left[ (\mathds{1} - \mathds{R}_2^\dagger \mathds{R}_2)  \mathds{D}^{12}(\mathds{1} - \mathds{R}_1 \mathds{R}_1^\dagger)  {\mathds{D}^{12}}^\dagger \right], & \kappa < \omega/c\\
     \Tr \left[ (\mathds{R}_2^\dagger - \mathds{R}_2) \mathds{D}^{12} (\mathds{R}_1 - \mathds{R}_1^\dagger)  {\mathds{D}^{12}}^\dagger \right]\re^{-2 |\gamma_{\rr}| d}  & \kappa > \omega/c ,
  \end{cases}
\end{split} 
\label{Eq:TransmissionCoeff}
\end{equation}
for propagating ($\kappa < \omega/c$) and evanescent ($\kappa > \omega/c$) modes 
where $\gamma_{\rr} = \sqrt{\omega^2/c^2 - \kappa^2}$, and $\mathds{D}^{12} = (\mathds{1} - \mathds{R}_1 \mathds{R}_2 \re^{2 \ri \gamma_{\rr} d})^{-1}$. 
The reflection matrix $\mathds{R}_i$ of the i-th structure is a 2x2 matrix in the polarization representation. Its four elements $R_{kl}$ with 
$k,l \in \{\rs, \rp\}$ for the scattering of s- or p-polarized plane waves into s- or p-polarized for the considered structures 
are determined with the method presented in Ref.~\cite{Rosa2008}. 

\begin{figure}
\begin{center}
\scalebox{0.35}{\includegraphics{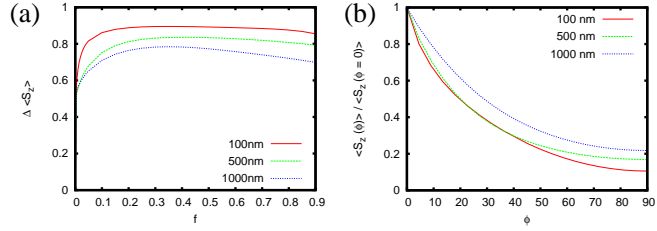}}
\caption{The heat flux between two Au gratings. (a) $\Delta \langle S_z \rangle(f)$ is defined as the difference in the heat flux between two parallel and two perpendicular gratings, normalized by the former (see also text).
(b) normalized by the flux $\langle S_z \rangle(0^\circ)$ when the gratings are aligned. The angle $\phi$ measures the relative twisting between the gratings, and the filling factor is fixed at $f=0.3$ }
\label{FigGoldFFactor}
\end{center}
\end{figure}

Now, we discuss the numerical results obtained with the above expressions for two Au gratings. In Fig.~\ref{FigGoldFFactor} (a) we show 
$\Delta\langle S_z \rangle \equiv [\langle S_z (\phi = 0^\circ) \rangle -  \langle S_z (\phi = 90^\circ) \rangle] / \langle S_z (\phi = 0^\circ) \rangle$ over the filling factor for three different distances $d = 100\,{\rm nm}$, $ 500\,{\rm nm}$ and $1000\,{\rm nm}$. Here, $\phi = \phi_1 - \phi_2$ is the twisting angle between the optical axes of the gratings. 
One can observe that the difference in the heat flux between both configurations, i.e.,
two parallel or perpendicular gratings, is larger than $70\%$ for all three distances and for a wide range of filling factors ranging from $f = 0.1$ to $0.9$ and can even reach about $90\%$ 
for $d = 100\,{\rm nm}$. 
In Fig.~\ref{FigGoldFFactor} (b) the dependence of $\langle S_z (\phi) \rangle$ on the twisting angle $\phi$ for a fixed filling factor of $f = 0.3$ shows that
for all considered distances the heat flux is very sensitive to a twisting of the gratings. The flux drops at least by $50\%$ for relatively small twisting angles of $\phi = 30^{\circ}$.
We note that $\langle S_z (\phi = 0) \rangle$ is about 12, 3.8, and 2.5 times the 
black body value ($\langle S_{\rm bb} \rangle \approx 459.3 {\rm W}/{\rm m}^2$) for $d = 100\,{\rm nm}$, $ 500\,{\rm nm}$, and $1000\,{\rm nm}$ and $f = 0.3$.

The observed sensitivity of the heat flux with respect to the twisting angle $\phi$ can be understood by the following consideration. When we have two parallel Au gratings, both still have a metallic response for a plane wave with polarization parallel to the grating structure. On the other hand, 
for a plane wave perpendicularly polarized to the gratings we have, according to Eq.~(\ref{Eq:EpsEMT}), a dielectric-like response dictated by $\epsilon_{\bot}$~\cite{DielectricResp}. For  
plane waves with polarizations between these two cases we have a mixture of metallic and dielectric response, but the crucial point is that we have always a symmetric situation which 
favors a larger heat flux \cite{asym}. By twisting the two Au gratings we break this symmetry. In particular, it is impossible to have a simultaneous metallic response of both gratings for any fixed polarization.

\begin{figure}
\begin{center}
\scalebox{0.35}{\includegraphics{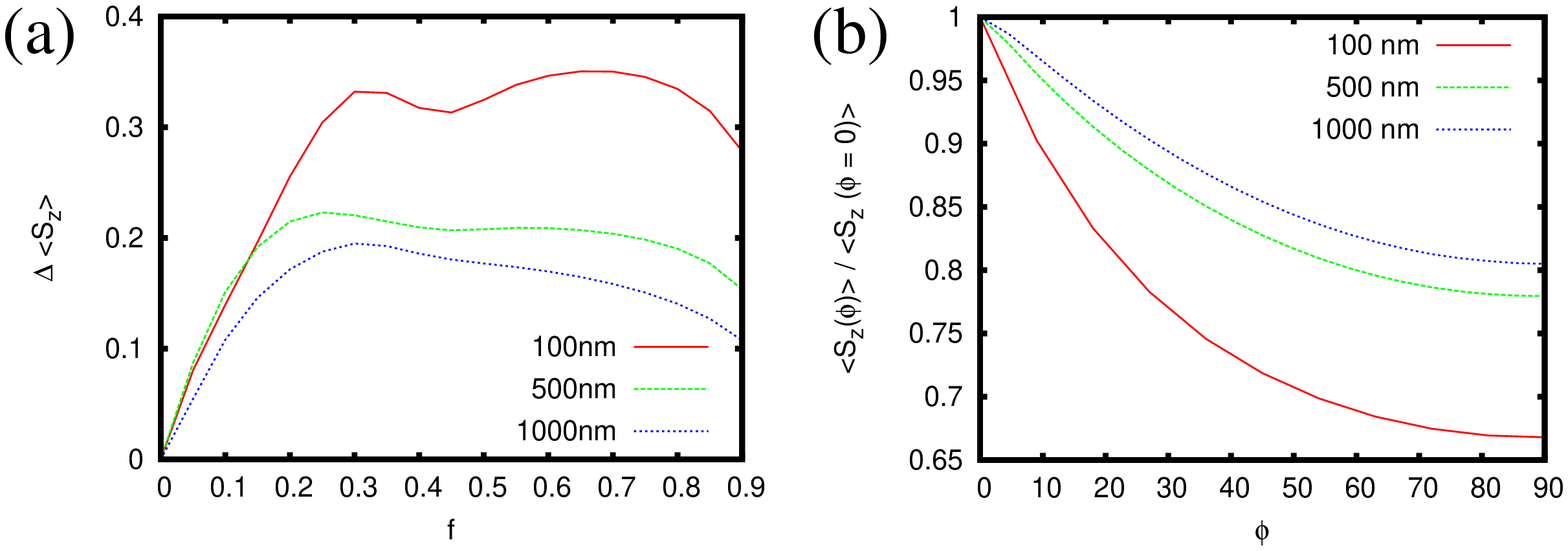}}
\caption{The same plots as in Fig.~\ref{FigGoldFFactor}, but for SiC instead of Au.}
\label{FigSiCFFactor}
\end{center}
\end{figure}

Finally, we will focus on the numerical results for two dielectric gratings made of SiC. In Fig.~\ref{FigSiCFFactor} (a) we show the same plots of $\Delta\langle S_z \rangle $ as in Fig.~\ref{FigGoldFFactor} (a)
but for two SiC gratings. Obviously, in this case the difference in the heat flux in the parallel and perpendicular configuration is smaller than for the two gold gratings and
it varies much more with respect to the filling factor. The heat flux is in this case also less sensitive with respect to the twisting angle as shown in Fig.~\ref{FigSiCFFactor} (b). 
We note that $\langle S_z (\phi = 0^\circ) \rangle$ is about 22.9, 4.2, and 2.8 times the black body value for respectively $d = 100\,{\rm nm}$, $ 500\,{\rm nm}$, and $1000\,{\rm nm}$, with $f = 0.3$.
While the drop in the heat flux when twisting the gratings can still be ascribed to the breaking of the symmetry, the underlying physical mechanisms are more involved, since in contrast to Au gratings for SiC gratings coupled surface modes and frustrated modes determine the heat flux at $T = 300\,{\rm K}$. By changing the filling factor the mode structure of these surface and frustrated modes
is changed and we find that new surface modes and bands of frustrated modes appear as was also found for porous media~\cite{BiehsEtAl2011}. By twisting the gratings the coupling between these modes becomes less efficient resulting in a smaller heat flux. 

In conclusion, we have theoretically shown that the near-field heat flux exchanged between two parallel grating structures can be modulated up to
$90\%$ by acting on the relative position of the optical axes of the gratings. We have also demonstrated that the flux magnitude is very sensitive to the
twisting angle, even for a low filling factors. This allows for manipulating the heat flux at nanoscale which can be usefull for thermal management in
micro/nano electromechanical devices. On the other hand, an efficient active control of transmission properties by mechanically driven grating structures
might be impractical. However, one can expect a similar effect for materials such as liquid crystals
or metal ferromagnetic structures for which the optical axis can be easily controled by applying external fields.

\begin{acknowledgments}
S.-A.\ B.\ gratefully acknowledges support from the Deutsche Akademie der Naturforscher Leopoldina
(Grant No.\ LPDS 2009-7). This research was partially supported by Triangle de la Physique, under
the contract 2010-037T-EIEM.
\end{acknowledgments}

%
%%%%%%%%%%%%%%%%%%%%%%%%%%%%%%%%%%%%%%%%%%%%%
%%%%%%%%%%%%%%%%%%%%%%%%%%%%%%%%%%%%%%%

%
%%%%%%%%%%%%%%%%%%%%%%%%%%%%%%%%%%%%%%%%%%%%%
%%%%%%%%%%%%%%%%%%%%%%%%%%%%%%%%%%%%%%%

%
%%%%%%%%%%%%%%%%%%%%%%%%%%%%%%%%%%%%%%%%%%%%%
%%%%%%%%%%%%%%%%%%%%%%%%%%%%%%%%%%%%%%%


\begin{thebibliography}{99}

% Intro
\bibitem{Polder1973} D. Polder and M. Van Hove, Phys. Rev. B {\bf 4} 3303 (1971).  

% Measurement of heat flux
\bibitem{Kittel} A. Kittel, W. M\"uller-Hirsch, J. Parisi, S.-A. Biehs, D. Reddig, and M. Holthaus, Phys. Rev. Lett. \textbf{95}, 24301 (2005).
\bibitem{HuEtAl2008} L. Hu, A. Narayanaswamy, X. Chen, and G. Chen, Appl. Phys. Lett. {\bf 92}, 133106 (2008).
\bibitem{NarayaEtAl2008} A. Narayanaswamy, S. Shen, and G. Chen, Phys. Rev. B {\bf 78}, 115303 (2008).
\bibitem{ShenEtAl2008} S. Shen, A. Narayanaswamy, and G. Chen, Nano Lett. {\bf 9}, 2909 (2009).

\bibitem{RousseauEtAl2009} E. Rousseau, A. Siria, G. Jourdan, S. Volz, F. Comin, J. Chevrier, and J.-J. Greffet, Nature Photonics {\bf 3}, 514 (2009).
\bibitem{Ottens2011} R. Ottens, V. Quetschke, S. Wise, A. Alemi, R. Lundock, G. Mueller, D. H. Reitze, D. B. Tanner, B. F. Whiting, preprint arXiv:1103.2389 (2011).

%layered media
\bibitem{Biehs2007} S.-A. Biehs, Eur. Phys. J. B {\bf 58}, 423-431 (2007).
\bibitem{PBA2009} P. Ben-Abdallah, Karl Joulain, J. Drevillon, and G. Domingues, J. Appl. Phys. {\bf 106}, 044306 (2009). 
%photonic crystals
\bibitem{PBA2010} P. Ben-Abdallah, K. Joulain, and A. Pryamikov. Appl. Phys. Lett {\bf 96}, 143117 (2010).
%meta materials
\bibitem{Joulain2010} K. Joulain, P. Ben-Abdallah, J. Drevillon, Phys. Rev. B {\bf 81}, 165119 (2010).
%pourous media
\bibitem{BiehsEtAl2011} S.-A. Biehs, P. Ben-Abdallah, F. S. S. Rosa, K. Joulain, and J.-J. Greffet, preprint arXiv:1103.2361v1 (2011).

%dipole model
\bibitem{OteyFan2011} C. Otey and S. Fan, preprint arXiv:1103.2668 (2011).
\bibitem{Krueger2011} Matthias Kr\"uger, Thorsten Emig, Mehran Kardar, preprint arXiv:1102.3891 (2011).

%spheroidal particles
\bibitem{HuthEtAl2010} O. Huth, F. R\"uting, S.-A. Biehs, and M. Holthaus, Eur. Phys. J. Appl. Phys. {\bf 50}, 10603 (2010).

% Metallic nanoparticle
\bibitem{ChapuisEtAl2008} P.-O. Chapuis, M. Laroche, S. Volz, and J.-J. Greffet, Phys. Rev. B {\bf 77}, 125402 (2008).
\bibitem{Nara2008} A. Narayanaswamy and G. Chen, Phys. Rev. B {\bf 77}, 075125 (2008). 
\bibitem{Domingues2005} G. Domingues, S. Volz, K. Joulain, and J.-J. Greffet, Phys. Rev. Lett. {\bf 94}, 085901 (2005).
\bibitem{Perez2008} A. P\'erez-Madrid, J. M. Rub\'i, and L. C. Lapas, Phys. Rev. B {\bf 77}, 155417 (2008).
\bibitem{Perez2009} A. P\'erez-Madrid, L. C. Lapas, and  J. M. Rub\'i, Phys. Rev. Lett. {\bf 103}, 048301 (2009).

% Review Volokitin and Persson
%\bibitem{VolokitinPersson2007} A. I. Volokitin and B. N. J. Persson, Rev. Mod. Phys. {\bf 79}, 1291 (2007).

%thermal imaging of structured surfaces
\bibitem{BiehsEtAl2008} S.-A. Biehs, O. Huth, and F. R\"uting, Phys. Rev. B {\bf 78}, 085414 (2008).
\bibitem{Kittel2008} A. Kittel, U. Wischnath, J. Welker, O. Huth, F. R\"uting, and S.-A. Biehs, Appl. Phys. Lett. {\bf 93}, 193109 (2008).
\bibitem{Rueting2010} F. R\"uting, S.-A. Biehs, O. Huth, and M. Holthaus, Phys. Rev. B {\bf 82}, 115443 (2010).
\bibitem{Biehs2010E} S.-A. Biehs, O. Huth, F. R\"uting, and M. Holthaus, J. Appl. Phys. {\bf 108}, 014312 (2010).

%Bimonte
\bibitem{Bimonte2009} G. Bimonte, Phys. Rev. A {\bf 80}, 042102 (2009).
\bibitem{Messina2010} R. Messina and M. Antezza, preprint arXiv:1012.5183 (2011).

%ladauer
\bibitem{BiehsEtAl2010} S.-A. Biehs, E. Rousseau, and J.-J. Greffet, Phys. Rev. Lett. {\bf 105}, 234301 (2010).
\bibitem{JoulainPBA2010} P. Ben-Abdallah and K. Joulain, Phys. Rev. B(R)  {\bf 82}, 121419 (2010).

%Phonon
\bibitem{ChangEtAl2006} C. W. Chang, D. Okawa, A. Majumdar, and A. Zettl, Science {\bf 314}, 1121 (2006). 
\bibitem{Wang2007} L. Wang and B. Li , Phys. Rev. Lett. {\bf 99}, 177208 (2007).
\bibitem{Segal2008} D. Segal, Phys. Rev. Lett. {\bf 100}, 105901 (2008).


%thermal rectification
\bibitem{Fan2010} C. R. Otey, W. T. Lau, and S. Fan, Phys. Rev. Lett. {\bf 104} 154301 (2010).
\bibitem{Zwol2010} P. J. van Zwol, K. Joulain, P. Ben-Abdallah, J.-J. Greffet, J. Chevrier, preprint arXiv:arXiv:1104.2994 (2011).


\bibitem{Mccauley10} Note that a related effect was observed for Casimir forces between anisotropic structures, see A. P. McCauley, F. S. S. Rosa, A. W. Rodriguez, J. D. Joannopoulos, D. A. R. Dalvit, S. G. Johnson , preprint arXiv:1009.4508v1 (2010).


% stoch. electrodynamics
\bibitem{Rytov} S. M. Rytov, Y. A. Kravtsov, and V. I. Tatarskii,
        {\em Principles of Statistical Radiophysics\/}, Vol.~3
        (Springer, New York, 1989).

% Effective Medium 
\bibitem{TaoEtAl1990} R. Tao, Z. Chen, and P. Sheng, Phys. Rev. B {\bf 41}, 2417 (1990). 
\bibitem{HaggansEtAl1993} C. W. Haggans, Lifeng Li, and R. K. Kostuk, J. Opt. Soc. Am. A {\bf 10}, 2217 (1993).
\bibitem{BrundrettEtAl1994} D. L. Brundrett, E. N. Glytsis, and T. K. Gaylord, Appl. Opt. {\bf 33}, 2695 (1994).

% anisotropy
\bibitem{Thesis} Habib Taouk, {\em Wave Propagation in General Anisotropic Media}, thesis (1986).
\bibitem{Rosa2008} F.S.S. Rosa, D.A.R. Dalvit, and P.W. Milonni, Phys. Rev. A {\bf 78} 032117 (2008).

\bibitem{DielectricResp} In the infrared $\epsilon_h$ is negative for metals and $|\epsilon_h| \gg 1$. Therefore, in this case even for small
filling factors $\epsilon_\perp > 0$ so that the response is not metallic anymore. From a physical point of view this is clear, since
by ruling the grating the conductivity in direction of the grating vector is inhibited. 

\bibitem{asym} We have carried out calculations between two isotropic semi-infinite Au bodies using $\epsilon_{1,2} = \epsilon_\parallel$ as permittivity.
We have compared these results with one semi-infinite medium with $\epsilon_1 = \epsilon_\parallel$ and the other with $\epsilon_2 = \epsilon_\perp$.
In the latter asymmetric case the heat flux was much smaller than the one found in the symmetric situation.

\end{thebibliography}
\end{document}